\documentclass[aps,reprint,amsmath,amssymb,superscriptaddress,showpacs,prb]{revtex4-1}
\usepackage{graphicx}
\usepackage{color}
\usepackage{lipsum}
\usepackage{dcolumn}
\begin{document}

\title{Parametrization of the Charge-Carrier Mobility in Organic Disordered Semiconductors. APAE against EGDM.}

\author {S.~D.~Baranovskii}
\affiliation{Department of Physics and Material Sciences Center,
Philipps-University, D-35032 Marburg, Germany}
\affiliation{Department f\"{u}r Chemie, Universit\"{a}t zu K\"{o}ln, Luxemburger Strasse 116, 50939 K\"{o}ln, Germany}

\author{A.~V.~Nenashev}
\thanks{On leave of absence from Rzhanov Institute of Semiconductor Physics
  and the Novosibirsk State University, Russia.}
\affiliation{Department of Physics and Material Sciences Center,
Philipps-University, D-35032 Marburg, Germany}


\author{D.~Hertel}
\affiliation{Department f\"{u}r Chemie, Universit\"{a}t zu K\"{o}ln, Luxemburger Strasse 116, 50939 K\"{o}ln, Germany}

\author{K.~Meerholz}
\affiliation{Department f\"{u}r Chemie, Universit\"{a}t zu K\"{o}ln, Luxemburger Strasse 116, 50939 K\"{o}ln, Germany}

\author{F.~Gebhard}
\affiliation{Department of Physics and Material Sciences Center,
Philipps-University, D-35032 Marburg, Germany}
\date{\today}

\date{\today}

\begin{abstract}
An appropriately parameterized compact analytical equation (APAE) is suggested to account for charge carrier mobility in organic disordered semiconductors (ODSs). This equation correctly reproduces the effects of temperature $T$, carrier concentration $n$, and electric field $F$ on the carrier mobility $\mu(T,F,n)$, as evidenced by comparison with analytical theories and Monte Carlo simulations. The set of material parameters responsible for charge transport is proven to be at varience to those used in the so-called extended Gaussian disorder model (EGDM) approach, which is widely exploited in commercially distributed device--simulation algorithms. While EGDM is only valid for cubic lattices with a specific choice of parameters, APAE describes charge transport in systems with spatial disorder in a wide range of parameters. APAE is user-friendly and, thus, suitable for incorporation into device-simulation algorithms.

  \end{abstract}







\maketitle   

\section{Introduction}
\label{introduction}
A correct parametrization of carrier mobility in organic disordered semiconductors (ODSs) with hopping charge transport is of vital importance for the development of algorithms desired to simulate devices based on ODSs, such as organic light–emitting diodes, organic solar cells, and organic field-effect transistors. Theoretical equations for the charge carrier mobility $\mu(T,n,F)$, dependent on the concentration of carriers $n$, on temperature $T$, and on the applied electric field $F$, are at the heart of the device simulation algorithms. The choice of the appropriate theoretical description for $\mu(T,n,F)$ suitable for applications in the device simulation software has been recently addressed in several studies  \cite{Kemerink2019,Lee2021,Sun2021,Oelerich2012,Nenashev2017short,Nenashev2017RANDOMLATTICE}. In spite of the progress achieved in these studies, the parametrization of the dependences $\mu(T,n,F)$ still needs improvement.

So far, most of the device simulation algorithms, including commercially distributed software packages, are based on the so-called EGDM equation \cite{Pasveer2005}. This equation has been suggested~\cite{Pasveer2005} as a ``unified description of charge-carrier mobilities in disordered semiconducting polymers''. Several review papers~\cite{Coehoorn_Bobbert_Feature2012,Kuik2014}
 promote the EGDM equation as gold standard for the description of carrier mobilities in ODSs, and the EGDM is the basis of commercially distributed device-simulation packages~\cite{Coehoorn_Bobbert_Feature2012,Lee2021}.

However, the EGDM equation~\cite{Pasveer2005} opposes the basic theoretical concepts developed so far for hopping transport in disordered materials~\cite{Mott1969,Shklovskii1973,Mott_Lecture_2018}. First, the EGDM equation is based on an irrelevant parametrization~\cite{Mott1969,Shklovskii1973,Mott_Lecture_2018}. Some parameters responsible for $\mu(T,n,F)$ are missed in the EGDM, while some parameters present in the EGDM are not responsible for the effects. Second, the EGDM equation was formulated~\cite{Pasveer2005} to fit simulation data on regular cubic lattices without spatial disorder.
However, hopping mobility on regular lattices is known to deviate significantly from that in materials with spatial disorder~\cite{Nenashev2017RANDOMLATTICE,Mott_Lecture_2018}.

Therefore, it is necessary to figure out whether the EGDM based on regular latices with deficient parametrization could be of use for systems with spatial disorder.

Regular cubic lattice used in EGDM is a specific case of the Gaussian-disorder-model (GDM), in which charge transport is due to incoherent
hopping of carriers via random spatially distributed localized states with Gaussian energy spectrum. This model seems valid by its ability to account for experimentally observed dependences $\mu(T,n,F)$~\cite{Bassler1993,Schmechel2002,Tessler2009}. A parametrization of the carrier mobility $\mu(T,n,F)$ in the framework of GDM is given in Sec.~\ref{sec:parametrization_GDM}.

While a lot of effort has been focused on computer simulations, the transport problem in the GDM can be easily solved analytically in the form of a simple closed-form system of equations \cite{Baranovskii2000,Baranovskii2002a,Rubel2004,Baranovski2006,Oelerich2012,Baranovskii2014,Nenashev_Topical_2015,Mott_Lecture_2018}. This solution is formulated in Sec.~\ref{sec:analytical_description}. Although this simple closed-form system of equations can easily be solved numerically, it has not yet become a state of the art for the device simulation community. Single user-friendly equations, like the EGDM, can be easier implemented into device-simulation software than systems of interconnected analytical equations, even so the user-friendliness is achieved at the cost of accuracy.

The challenging task is to replace the system of analytical equations for $\mu(T,n,F)$ by a single user-friendly appropriately parameterized analytical equation (APAE) that can be easily embedded into device-simulation software.
Our APAE is formulated in Sec.~\ref{sec:APAE}, providing the main result of our paper.

In Sec.~\ref{sec:proofs_APAE}, the validity of the APAE is proven by comparison with the results of computer simulations. The agreement of this heuristic equation with analytical theories and with computer simulations suggests the APAE for using in the device-simulation algorithms.

In Sec.~\ref{sec:failure_of_EGDM}, the widely used alternative description of $\mu(T,n,F)$ based on the EGDM is discussed. In spite of the insufficient parametrization and of the reduction of simulated systems to regular lattices, the EGDM could be applicable to some systems with spatial disorder.  Sec.~\ref{sec:discussion} is dedicated to comparison with recent studies in the literature, highlighting the advantages of the APAE.

\section{Crucial parameters for the carrier mobility}
\label{sec:parametrization_GDM}

\subsection{Gaussian Disorder Model (GDM)}
\label{sec:GDM}
It has been established that charge transport in single-component and multicomponent ODSs is due to incoherent hopping of carriers via randomly distributed localized states with Gaussian energy spectrum~\cite{Silinsh1970,Bassler1993,Pasveer2005,Baranovski2006,Mensfoort2008,Germs2011,Oelerich2012,Coehoorn_Bobbert_Feature2012,Kuik2014,Baranovskii2014,Nenashev_Topical_2015,Mott_Lecture_2018,Kemerink2019,Sun2021,Lee2021}, 
\begin{equation}
\label{DOS_Gauss}
g(\varepsilon) = \frac{N}{\sigma\sqrt{2\pi}}\exp\left(-\frac{\varepsilon^{2}}{2\sigma^{2}}\right).
\end{equation}
Here, $\sigma$ is the energy scale of the density of states (DOS) and $N$ is the concentration of randomly distributed localized states, called henceforth hopping ``sites''. The estimates for $\sigma$ vary between $\sigma \simeq 0.05$ eV and $\sigma \simeq 0.15$ eV \cite{Bassler1993,Pasveer2005,Kemerink2019}, the estimates for $N$ vary between $N \simeq 1.7 \times 10^{20}$ cm$^{-3}$ and $N \simeq 4.6 \times 10^{21}$ cm$^{-3}$, depending on the material \cite{Bassler1993,Pasveer2005,Lee2021}. For the sake of simplicity, we follow most previous studies considering the GDM without correlations between the spatial positions of hopping sites and their energies \cite{Bassler1993,Pasveer2005,Baranovski2006,Mensfoort2008,Germs2011,Oelerich2012,Coehoorn_Bobbert_Feature2012,Kuik2014,Baranovskii2014,Nenashev_Topical_2015,Mott_Lecture_2018,Kemerink2019,Sun2021,Lee2021}.

Phonon-assisted hopping suggested by Miller and Abrahams \cite{Miller1960} is usually considered as the dominant charge transport mechanism in ODSs~\cite{Bassler1993,Pasveer2005,Baranovski2006,Mensfoort2008,Germs2011,Oelerich2012,Coehoorn_Bobbert_Feature2012,Kuik2014,Baranovskii2014,Nenashev_Topical_2015,Mott_Lecture_2018,Kemerink2019,Sun2021,Lee2021}. The expression for the rate of carrier transfer from the occupied site with energy $\varepsilon_i$ to the empty site with energy $\varepsilon_j$ over the distance $r_{ij}$ has the form

\begin{equation}
	\label{eq:nu-ij}
	\nu_{ij} = \nu_0 \exp\left( -\frac{2|\mathbf{r}_{ij}|}{a} \right) \chi\left(\frac{\varepsilon_j-\varepsilon_i-e\mathbf{F}\cdot\mathbf{r}_{ij}}{kT}\right)
\end{equation}
with
\begin{equation}
\nonumber
	\label{eq:gamma}
	\chi(X) =
	\begin{cases}
		\exp(-X) , & \text{if $X>0$,} \\
		1,                      & \text{if $X\leq0$}  .
	\end{cases}
\end{equation}
Here, $\mathbf{F}$ is the applied electric field, $e$ is the elementary charge, $a$ is the localization length of charge carriers in the localized states,
$k$ is the Boltzmann constant,
and $T$ is
temperature. Estimates of $a$ in the range $0.1 \leq a \leq 0.75$ nm have been suggested in the literature \cite{Gill1972,Rubel2004,Pasveer2005,Lee2021}. The prefactor in Eq.~(\ref{eq:nu-ij}) is usually described by a single parameter, the so-called attempt-to-escape frequency $\nu_0$. Precise quantum-mechanical calculation of $\nu_0$ can be found elsewhere \cite{Miller1960,Shklovskii1984}.
The energy $\varepsilon_i$ of the starting site and the energy $\varepsilon_j$ of the target site in Eq.~(\ref{eq:nu-ij}) are counted without contributions of the applied electric field. The effect of the electric field $\mathbf{F}$ on the hopping rates is expressed explicitly by the term $e\mathbf{F}\cdot\mathbf{r}_{ij}$ in the exponent of the r.h.s. in Eq.~(\ref{eq:nu-ij}).

The validity of the GDM determined by Eqs.~(\ref{DOS_Gauss}) and (\ref{eq:nu-ij}) is justified by its ability to account for a broad variety of experimental observations. Among those is the transition from the dependence $\ln[\mu(T)] \propto 1/T^2$ to the dependence $\ln[\mu(T)] \propto 1/T$ with rising carrier concentration\cite{Baranovskii2002a,Baranovski2006,Baranovskii2014}. Another pronounced phenomenon predicted by Eqs.~(\ref{DOS_Gauss}) and (\ref{eq:nu-ij}) is the transition from the mobility $\mu$ independent on carrier concentration $n$ at small $n$ to the mobility $\mu$ strongly dependent on $n$ at large $n$ values  \cite{Baranovskii2002a,Baranovski2006,Oelerich2012,Baranovskii2014,Nenashev_Topical_2015,Mott_Lecture_2018}. Experimental data to the latter effect~\cite{Tanase2003} are indicative~\cite{Oelerich2012,Baranovskii2014,Nenashev_Topical_2015,Mott_Lecture_2018} for the Gaussian shape of the DOS given by Eq.~(\ref{DOS_Gauss}). Therefore, our consideration is based on Eqs.~(\ref{DOS_Gauss}) and (\ref{eq:nu-ij}), in agreement with several recent studies~\cite{Mott_Lecture_2018,Kemerink2019,Sun2021,Lee2021}.

\subsection{Parametrization of the mobility $\mu(T,n,F)$}
\label{sec:parametrizzation}

For the discusion of the proper parametrization of $\mu(T,n,F)$, we focus on strong exponential dependencies of the carrier mobility $\mu(T,n,F)$, namely, on temperature $T$, electric field $F$, and carrier concentration $n$.

\subsubsection{Parametrization at small electric fields}
\label{sec:small_F_parametrization}

As evident from the exponents in Eqs.~(\ref{DOS_Gauss}), (\ref{eq:nu-ij}), the transport problem at small electric fields, $F \rightarrow 0$, is determined by only two dimensionless parameters: $\widetilde{\alpha} = kT/\sigma$ and $\beta = N^{-1/3}/a$. This is true at low carrier concentrations~\cite{Baranovskii2000,Baranovskii2002a,Rubel2004}, $n \ll N$, when the mobility $\mu$ does not depend on $n$. At elevated $n$, a third dimensionless parameter, $\gamma = n/N$, enters the carrier mobility $\mu(T,n)$.

Remarkably, the dependence of the carrier mobility $\mu$ on temperature $T$ is affected by the parameter~ ($N^{-1/3}/a$).\cite{Mott1969}
This is true in the so-called variable-range hopping (VRH) regime, when the characteristic hopping length depends on $T$. Charge transport in ODSs is dominated by the VRH process, as has been proven by Monte Carlo simulations and analytical calculations~\cite{Baranovskii2000,Baranovskii2002a,Rubel2004,Nenashev2017short,Nenashev2017RANDOMLATTICE}. Therefore, the temperature dependence $\mu(T,n)$ is sensitive to $\beta$. \cite{Baranovskii2000,Baranovskii2002a,Rubel2004,Nenashev2017short,Nenashev2017RANDOMLATTICE} This effect is often overlooked~\cite{Bassler1993,Pasveer2005}.


\subsubsection{Parametrization of the field dependence}
\label{sec:large_F_parametrization}

The appropriate parametrization of the dependence $\mu(F)$ in hopping transport has been revealed in 1973 by Shklovskii~\cite{Shklovskii1973}, who concluded that the effect of the electric field $F$ on the carrier mobility $\mu$ is determined by the product $eaF$, where $a$ is the
localization length. Shklovskii considered for simplicity the case $T=0$, recognizing that a charge carrier gains the amount of energy $\Delta = eFx$ tunnelling in the field direction over some distance $x$. The tunneling rate $\nu(x) \propto \exp(-2x/a)$ can be rewritten as $\nu(x) \propto \exp(-\Delta/kT_{\mathrm{eff}})$ with $T_{\mathrm{eff}} \simeq eFa/2$. Apparently, the field-dependent effective temperature $T_{\mathrm{eff}} \simeq eFa/2$ accounts for the effect of electric field $F$ on hopping transport at $T=0$.

For the case $T \neq 0$, Marianer and Shklovskii~\cite{Marianer1992} suggested that the combined effects of electric field $F$ and temperature $T$ can be expressed in the form of the effective temperature

\begin {equation}
\label {eq:Teff}
T_\text{eff} = T \left[1 + \left(c_1
\frac{e F a} {kT} \right)^2 \right]^{1/2}
\end {equation}
with $c_1 \approx 0.67$. Several studies performed by numerical simulations~\cite{Hess1993,Cleve1995,Jansson2008PRB,Nenashev2017short} confirmed the validity of this approach with values $c_1$ distributed in the range $0.5 \leq c_1 \leq 0.9$.

Apparently, the localization length $a$ and not the intersite distance $N^{-1/3}$ governs the effect of electric field on the hopping conductivity. This fact is non-trivial because the electric field enters the theory only via the combination $e\mathbf{F}\cdot\mathbf{r}_{ij}$, in which the length of a hop $|\mathbf{r}_{ij}|$ is of the order of the intersite distance $N^{-1/3}$. Therefore, one might expect the combination of parameters $e N^{-1/3} F$ to be essential for the field-dependent mobility. However, it has been rigorously proven by straightforward computer simulations~\cite{Nenashev2017short} that the localization length $a$, i.e., the feature of a single localized state, and not the intersite distance $N^{-1/3}$, is responsible for $\mu(F)$. 
This counterintuitive result has not yet been adopted by the broad scientific community in spite of its rigorous proof~\cite{Nenashev2017short,Mott_Lecture_2018}. Equation (\ref{eq:Teff}) along with Eqs.~(\ref{DOS_Gauss}) and (\ref{eq:nu-ij}) implies that the combined effects of electric field $\mathbf{F}$ and temperature $T$ are described by a single parameter $\alpha = kT_\text{eff}/\sigma$. It means that the effect of the electric field on the hopping conductivity is governed in accord with Eq.~(\ref{eq:Teff}) by the parameter
\begin {equation}
\label{eq:delta}
\delta = eFa/(kT) \, .
\end {equation}

Herewith, only three parameters are responsible for $\mu(T,n,F)$, namely,
\begin {equation}
\label{eq:alpha_betta_gamma_delta}
\alpha = kT_\text{eff}/\sigma; \
\beta = N^{-1/3}/a; \ \gamma = n/N\ 
.
\end {equation}

\section{Analytical description of $\mu(T,n,F)$}
\label{sec:analytical_description}

The analytical theory for the description of hopping transport in amorphous materials with strongly energy-dependent DOS $g(\varepsilon)$ is known for decades~\cite{Shklovskii1990,Baranovskii2005,Baranovski2006}. In particular, it has been proven~\cite{Oelerich2012,Baranovskii2014,Nenashev_Topical_2015,Mott_Lecture_2018} that the charge carrier mobility can be desribed in the framework of the GDM as
\begin{align}
  \label{eq:mob}
  \begin{split}
  \mu =\mu_0 \gamma^{-1}
   \exp\left(-\frac{2B_{c}^{1/3}}{a}r(\varepsilon_t)
  -\frac{\varepsilon_t-\varepsilon_F}{kT}\right) \,,
  \end{split}
\end{align}
where
\begin{align}
  \label{eq:r_t}
  r(\varepsilon_t)= \left[\frac{4\pi}{3}\int\limits_{-\infty}^{\varepsilon_t}
   g(\varepsilon')
    [1-f(\varepsilon',\varepsilon_{F})] d\varepsilon' \right]^{-1/3} \,  ,
\end{align}
and the transport energy, $\varepsilon_t$, is calculated from~\cite{Oelerich2012,Baranovskii2014,Mott_Lecture_2018}
\begin{align}
  \label{eq:et}
  \begin{split}
    \frac{2}{3}\left(\frac{4\pi}{3B_{c}}\right)^{-\frac{1}{3}}
      \frac{kT}{a}
    & \left[\int_{-\infty}^{\varepsilon_{t}} [1-f(\varepsilon,\varepsilon_F)]
      g(\varepsilon)d\varepsilon\right]^{-\frac{4}{3}} \\
    &\times[1-f(\varepsilon_t,\varepsilon_F)] g(\varepsilon_t) = 1 \, ,
  \end{split}
\end{align}
where $f(\varepsilon,\varepsilon_{F})$ is Fermi function
\begin{align}
  \label{Fermi function}
  f(\varepsilon,\varepsilon_{F}) = \left[1 + \exp{\frac{(\varepsilon - \varepsilon_{F})}{kT}}\right]^{-1} \,
\end{align}
with the Fermi energy $\varepsilon_{F}$ determined by the relation
\begin{align}
  \label{eq:Fermi_MT}
  \int_{-\infty}^{\infty} g(\varepsilon)f(\varepsilon, \varepsilon_{F})d\varepsilon =n \, ,
\end{align}
which accounts for the finite charge carrier concentration $n$. The coefficient $B_{c} \simeq 2.7$ is due to the percolation nature of the hopping transport~\cite{Rubel2004}.

The preexponential factor in Eq.~(\ref{eq:mob}) has the value
\begin{align}
  \label{eq:mu_0}
 \mu_0 = B \frac{e\nu_0}{kTN^{2/3}} \, ,
\end{align}
where $B$ is a numerical factor, which can be determined by comparison with computer simulations.

The structure of prefactor $\mu_0$ in Eq.~(\ref{eq:mu_0}) relies on the conventional form of the Einstein relation between the carrier mobility and the carrier diffusion coefficient that is valid for non-degenerate systems, i.e., at low carrier concentrations $n \ll N$. In general, the ratio $e/kT$ should be replaced by the generalized Einstein relation~\cite{Roichman2002,Tessler2009,Baranovskii2014}. We nevertheless leave Eq.~(\ref{eq:mu_0}) in the given form because the final result for $\mu(T,n,F)$ in the next Sec.~\ref{sec:APAE} is obtained from a comparison of the parameters in the analytical theory with numerical data from computer simulations. In this respect, our approach is essentially similar to that exploited in the recent studies by Upreti \textit{et al}.~\cite{Kemerink2019} and by Lee et al.~\cite{Lee2021}, who calibrated parameters of a similar analytical theory by comparison with numerical simulations.

\section{Appropriately parameterized analytical equation (APAE) for $\mu(T,n,F)$}
\label{sec:APAE}

The set of Eqs.~(\ref{eq:mob})-(\ref{eq:mu_0}) determines $\mu(T,n,F)$ in terms of the relevant parameters given by Eq.~(\ref{eq:alpha_betta_gamma_delta}). Remarkably, the function $\mu(T,n,F)$ obtained from this set of equations can be approximated in the realistic parameter ranges by a single appropriately parameterized analytical equation (APAE)
\begin{align}
  \label{eq:APAE_general}
  \begin{split}
  \mu(T,n,F)= \frac{e\nu_0}{\sigma N^{2/3}}\Phi(\gamma) \exp\left[- A(\beta) -  \alpha^{-2}C(\beta) \right]
     \, ,
  \end{split}
\end{align}
where
\begin{align}
  \label{eq:APAE_A_betta}
 A(\beta) = c_2\beta^2+c_3\beta +c_4
     \, ,
\end{align}
\begin{align}
  \label{eq:APAE_C_betta}
 C(\beta) = c_5\beta^{-2}+c_6\beta^{-1} +c_7
     \, .
\end{align}
The constant $\exp(c_4)$ replaces the factor $B$ in Eq.~(\ref{eq:mu_0}).

The function $\Phi(\gamma)$ in Eq.~(\ref{eq:APAE_general}), responsible for the dependence $\mu(n)$, is determined by the equation
\begin{align}
  \label{eq:APAE_Phi}
  \Phi(\gamma) = \exp\left[(c_8\ln\alpha^{-1}-c_9)\xi^2 \Theta(\xi) \right] \, ,
\end{align}
where
\begin{align}
  \label{eq:APAE_xi}
  \xi = \frac{\alpha^{-1}}{2}+2-\left|\ln\left(\gamma-\frac{c_{10}\gamma^2}{\max(\beta,c_{11})} \right) \right|^{1/2} \, ,
\end{align}
\begin{equation}
\nonumber
	\label{eq:APAE_Theta}
	\Theta(x) =
	\begin{cases}
		1 , & \text{if $x\geq0$,} \\
		0,                      & \text{if $x<0$}  .
	\end{cases}
\end{equation}

Numerical parameters $c_1$ to $c_{11}$ were further optimized by comparison with computer simulations attaining the values
\begin{equation}
\begin{split}
c_1 &= 0.6\\
c_2 &= -0.066\\
c_3 &= 2.65\\
c_4 &= -1.35\\
c_5 &= 0.89\\
c_6 &= -0.86\\
c_7 &= 0.54\\
c_8 &= 0.6\\
c_9 &= 0.15\\
c_{10} &= 13\\
c_{11} &= 3
\end{split}
\label{eq:c_parameters}
\end{equation}

Recall that $\alpha = kT_\text{eff}/\sigma,
\beta = N^{-1/3}/a,  \gamma = n/N$, see Eq.~(\ref{eq:alpha_betta_gamma_delta}), and $T_{\text{eff}}$ is determined by Eq.~(\ref{eq:Teff}) with parameter $c_1$ given by Eq.~(\ref{eq:c_parameters}). Equation (\ref{eq:APAE_general}) is the central result of our work.

\section{Testing the APAE by computer simulations}
\label{sec:proofs_APAE}

\subsection{APAE and $\mu(F)$}
\label{sec:APAE_and_mu_F}

\begin{figure}
	\includegraphics[width=\linewidth]{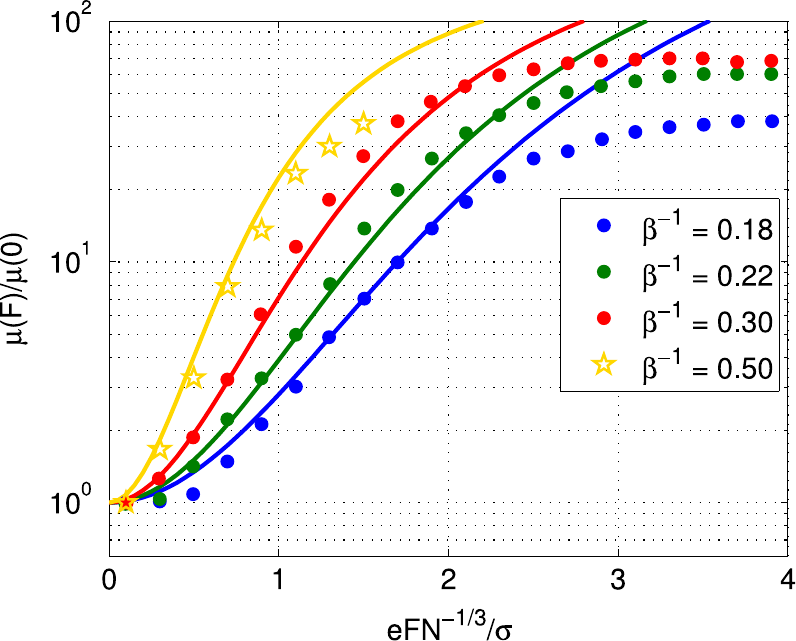}
	\caption{Normalized mobility $\mu(F)/\mu(0)$ at $\sigma/kT = 4$ and $n \rightarrow 0$ as a function of $eFN^{-1/3}/\sigma$ at different values of $\beta=N^{-1/3}/a$. Symbols: results of MC simulations; solid lines: results of the APAE equation (\ref{eq:APAE_general}).  }
	\label{fig:mu_F}
\end{figure}

In Fig.~\ref{fig:mu_F}, the dependences of the normalized mobility $\mu(F)/\mu(0)$ obtained by Monte Carlo (MC) simulations at different values of the parameter $\beta = N^{-1/3}/a$ are depicted by symbols. Simulations were carried out in the framework of the model with spatial and energy disorder formulated in Sec.~\ref{sec:GDM}. The simulation algorithm is the same as the one used in previous studies~\cite{Nenashev2017short}. In particular, we used the HopHop algorithm (https://github.com/janoliver/hophop). Simulation data at $\beta^{-1} \leq 0.3$ agree with those from the previous studies~\cite{Nenashev2017short}, while the data at $\beta = 2$ are novel. At all values of $\beta$ in Fig.~\ref{fig:mu_F}, $\mu(F)/\mu(0)$ steeply increases with rising field $F$ at $F\leq 2\sigma/ eN^{-1/3}$.
At larger fields, the field dependences in Fig.~\ref{fig:mu_F} saturate. Solid lines in Fig.~\ref{fig:mu_F} depict the results of the APAE. Apparently, APAE appropriately describes the simulation data at $F\leq 2\sigma/ eN^{-1/3}$, while at larger fields, APAE predicts a stronger dependence $\mu(F)/\mu(0)$ than the one yielded by simulations.

The weakening of the field dependency at high fields in Fig.~\ref{fig:mu_F} is not surprising. There are several mechanisms contributing to this effect~\cite{Baranovskii2014,Mott_Lecture_2018}. In fact, $\mu(F)$ in hopping transport should even decrease at large electric fields~\cite{NS,Levin1988,Nenashev2008NDC}. Such effects at very high electric fields cannot be described by the effective temperature in Eq.~(\ref{eq:Teff}).

It is seen in Fig.~\ref{fig:mu_F} that APAE based on Eq.~(\ref{eq:Teff}) describes the dependence $\mu(F)$ only at electric fields smaller than $F^* \simeq 2\sigma/ eN^{-1/3}$. Taking relevant parameter values for ODSs~\cite{Pasveer2005,Lee2021},
$N \approx 10^{21}$ cm$^{-3}$ and $\sigma \approx 0.1$ eV,
one obtains $F^* \approx 2 \times 10^6$ Vcm$^{-1}$. The range of electric fields $F < 2 \times 10^6$ Vcm$^{-1}$ is relevant to experimental studies~\cite{Hirao1995,Mozer2005} and to device applications~\cite{Kemerink2019}. Hence, the APAE can be used to interpret experimental data and it can be applied in device simulation algorithms.

\subsection{APAE and $\mu(T)$}
\label{sec:APAE_and_mu_T}

The scale of energy disorder $\sigma$ is usually estimated in ODSs between $\sigma \simeq 50$ meV and $\sigma \simeq 150$ meV \cite{Bassler1993,Kemerink2019,Lee2021}. This gives for the ratio $\sigma/(kT)$  at room temperature $T \thickapprox 300$ K the values between $\sigma/(kT) \simeq 2$ and $\sigma/(kT) \simeq 6$.

\begin{figure}
	\includegraphics[width=\linewidth]{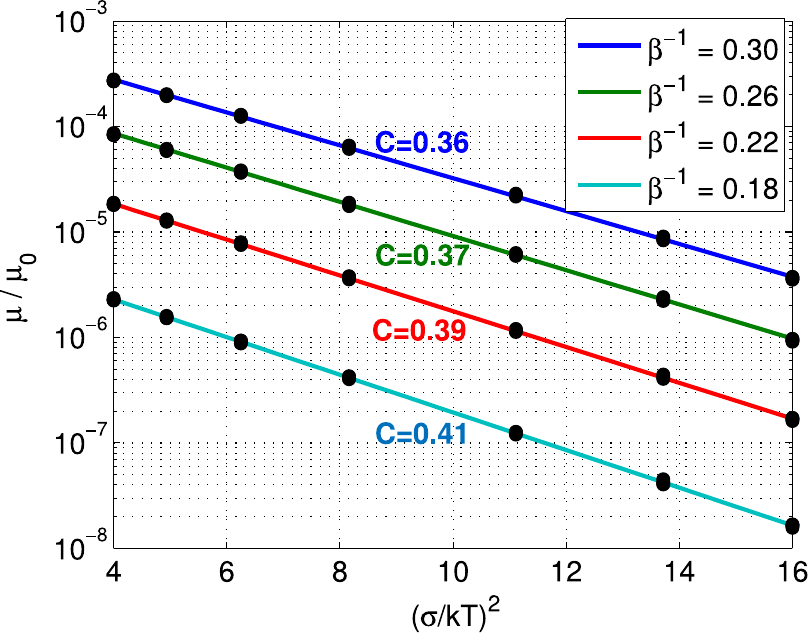}
	\caption{Mobility $\mu(T)$ at $n \rightarrow 0$, $F \rightarrow 0$ as a function of $\alpha^{-2} =  \sigma^2/(kT)^2$ for different values of $\beta = N^{-1/3}/a$. Symbols: results of simulations; solid lines: results of the APAE.}
	\label{fig:mu_low_n}
\end{figure}

In Fig.~\ref{fig:mu_low_n}, the mobility $\mu(F \rightarrow 0, n \rightarrow 0)$, as obtained by computer simulations, is plotted as a function of $\sigma/(kT$) at different values of $\beta = N^{-1/3}/a$.  The results of the APAE are in perfect agreement with the simulation data. In contrast to the MC simulations used for the data in Fig.~\ref{fig:mu_F}, numerical calculations at $F \rightarrow 0$ were performed by solution of Kirchhoff equations in a resistor network proposed by Miller and Abrahams~\cite{Shklovskii1984}.

The data in Fig.~\ref{fig:mu_low_n} correspond to the well-known dependence $\mu(T)$, which can be approximated by the expression
\begin{equation}
\mu(T) \propto  \exp\left[-C \left(\frac{\sigma}{kT}\right)^2\right] .
\label{eq:mu_T_Bassler}
\end{equation}
By fitting the Monte Carlo simulation data for $\mu(T)$ obtained on a cubic lattice at $\beta = N^{-1/3}/a =5$, B\"{a}ssler~\cite{Bassler1993} suggested the value $C=(2/3)^2 \simeq 0.44$. Using analytical theory based on the approach of transport energy given by Eq.~(\ref{eq:et}), it was shown~\cite{Baranovskii2000} that in the GDM with spatial and energy disorder described in Sec.~\ref{sec:GDM}, coefficient $C$ depends on parameter $\beta = N^{-1/3}/a$ having the values $C \simeq 0.46$ at $\beta =10$, $C \simeq 0.41$ at $\beta =5$ and $C \simeq 0.38$ at $\beta \simeq 3.7$. These values have been confirmed by analytical calculations based on the percolation theory~\cite{Baranovskii2002a}. The data depicted in Fig.~\ref{fig:mu_low_n} are in good agreement with these previous results yielding the values $C$ specified in the figure.

\subsection{APAE and $\mu(n)$}
\label{sec:APAE_and_mu_n}

\begin{figure}
	\includegraphics[width=\linewidth]{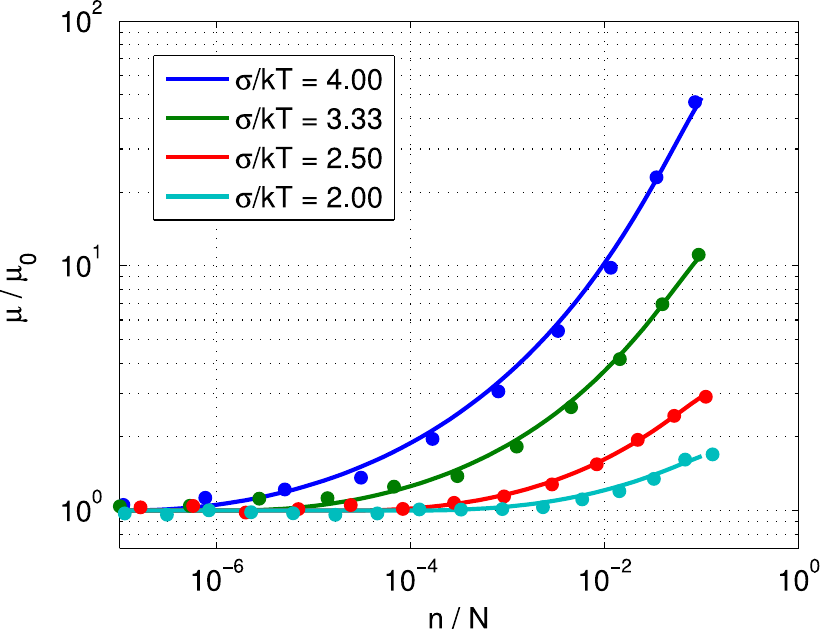}
	\caption{Normalized mobility $\mu(n)/\mu(n \rightarrow 0)$ at $F \rightarrow 0$ as a function of $\gamma = n/N$ at $\beta = 5$ and different values of $\alpha =kT/\sigma$. Symbols: results of simulations; solid lines: results of the APAE.}
	\label{fig:mu_n}
\end{figure}

In Fig.~\ref{fig:mu_n}, the normalized mobility $\mu(n)/\mu(n \rightarrow 0)$ at $F \rightarrow 0$ is plotted as a function of $\gamma = n/N$ at a typical ODSs parameter~\cite{Bassler1993} $\beta = 5$ and different values of $\alpha =kT/\sigma$. The choice of $\beta \simeq 5$ is further supported by the estimates~\cite{Rubel2004}  $0.1$ nm $\leq a \leq 0.3$ nm and  $0.5$ nm $\leq N^{-1/3} \leq 1.8$ nm in ODSs~\cite{Rubel2004,Pasveer2005}.

Apparently, the computer simulations confirm the capability of the APAE to describe the dependences $\mu(T,n,F)$ in the framework of the GDM, i.e., for hopping transport via randonly distributed localized states with Gaussian energy spectrum.

Most algorithms developed so far for the simulation of devices based on ODSs, such
as organic light–emitting diodes, organic solar cells, and organic field-effect transistors, are based on the EGDM equation~\cite{Pasveer2005,Coehoorn_Bobbert_Feature2012}. In the following sections, we analyze the EGDM equation and compare it with APAE.

\section{Comparison with EGDM}
\label{sec:failure_of_EGDM}

A reduced version of the GDM without spatial disorder is, in fact, what one calls EGDM~\cite{Coehoorn_Bobbert_Feature2012}. Charge transport on a regular cubic lattice was simulated for the case $b = 10a$, where $b$ is the lattice spacing and $a$ is the localization length in Eq.~(\ref{eq:nu-ij}). To make a bridge to the case of randomly placed localized states, we note that $b = N^{-1/3}$. Simulation results for the carrier mobility $\mu(T,n,F)$ were fitted in the form \cite{Pasveer2005}
\begin{align}
	\label{eq:ElectrFieldPasveer_1}
	\mu(T,n,F)\approx \widetilde{\mu}(T,n) \phi(T,F) \, ,
\end{align}
where $\phi(T,F)$ is
\begin{align}
	\label{eq:ElectrFieldPasveer}
	\begin{split}
	\phi(T,F) = \exp\Biggl\{
	  & 0.44 \left[\widehat{\sigma}^{3/2}-2.2\right]            \\
	  & \times \left[\sqrt{1+0.8\left(\frac{Feb}{\sigma}\right)^{2}}-1\right]
	\Biggr\} \, 
	\end{split}
\end{align}
with $\widehat{\sigma} = \sigma/(kT)$. The function $\widetilde{\mu}(T,n)$ is given by~\cite{Pasveer2005}
\begin{subequations}
  \begin{align}
  \widetilde{\mu}(T,n) = \mu_0(T)
 \exp[(\widehat{\sigma}^2-\widehat{\sigma})(2n b^3)^{\widetilde{\delta}} /2] ,  & \\
     \mu_0(T) = \mu_0 \mathfrak{c}_1 \exp[-\mathfrak{c}_2 \widehat{\sigma}^2] , & \label{eq:Pasveer_T_n_part_b} \\
        \widetilde{\delta} \equiv 2\frac{\ln(\widehat{\sigma}^2-\widehat{\sigma})-\ln(\ln 4)}{\widehat{\sigma}^2}, \,  \mu_0\equiv\frac{b^2 \nu_0 e}{\sigma}, &
 \label{eq:Pasveer_T_n}
 \end{align}
\end{subequations}
with $\mathfrak{c}_1 = 1.8 \times 10^{-9}$ and $\mathfrak{c}_2 = 0.42$.
Equations (\ref{eq:ElectrFieldPasveer_1}) -- (\ref{eq:Pasveer_T_n}) were named \cite{Coehoorn_Bobbert_Feature2012} the
extended Gaussian disorder model (EGDM). These equations fit numerical data for $\mu(T,n,F)$ obtained on a cubic lattice at $b/a=10$.

Remarkably, the localization length of charge carriers, $a$, does not enter the EGDM equations at all, although $a$ is known~\cite{Mott1969,Shklovskii1973} to determine the temperature- and the field-dependencies of the hopping mobility via parameters $\alpha = kT_\text{eff}/\sigma$,  $\beta = N^{-1/3}/a$ present in Eqs.~(\ref{eq:Teff}), (\ref{eq:alpha_betta_gamma_delta}).

The absence of $a$ in the EGDM, was supported by two arguments.
First, it was stated that presumably in all published 3D-modeling work on the (E)GDM a fixed value $N^{-1/3}/a=10$ had been used, suggesting $N^{-1/3}/a=10$ as a standard one for ODSs~\cite{Coehoorn_Bobbert_Feature2012}. In this respect, one has to remark, however, that published 3D-modeling work on the GDM had mostly used a fixed value $N^{-1/3}/a=5$ instead~\cite{Bassler1993}.

Second, it was stated, referring to unpublished work, that varying the ratio $N^{-1/3}/a$ has no significant effect on the temperature, field, or carrier density dependence of the mobility~\cite{Coehoorn_Bobbert_Feature2012}.
Fig.~\ref{fig:mu_F} evidences, however, that varying the ratio $N^{-1/3}/a$ does have a significant effect on the field dependence of the mobility in the GDM with spatial disorder. A similar effect on cubic lattices has been proven elsewhere~\cite{Nenashev2017short,Mott_Lecture_2018}.

This analysis raises a question of whether EGDM can be used for description of $\mu(T,n,F)$ in ODSs. The answer to this question is of high importance not only for academic researchers, but particularly for the community dealing with device simulations, where EGDM is, so far, considered as the state of the art.
It is a lucky coincidence that the EGDM equation could be, in some cases, applicable to ODSs.

\begin{figure}
	\includegraphics[width=\linewidth]{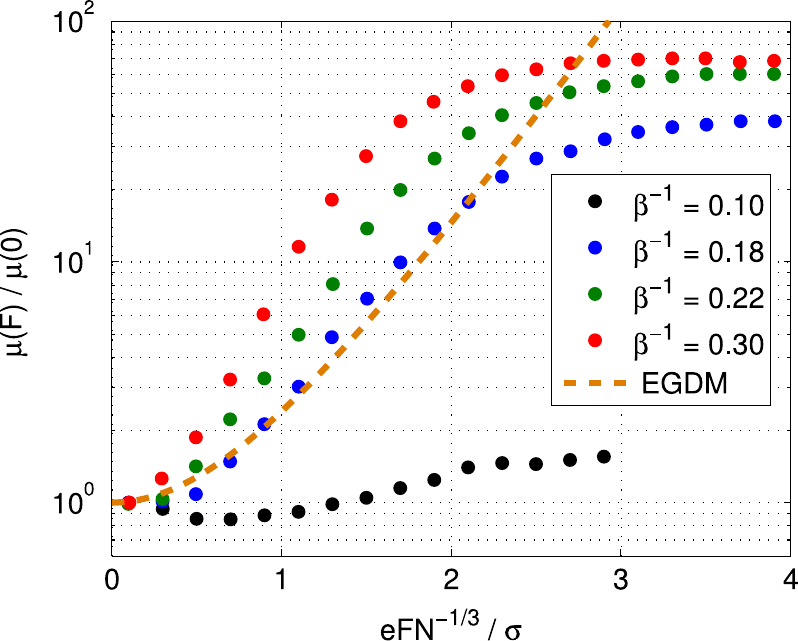}
	\caption{Normalized mobility $\mu(F)/\mu(0)$ at $n \rightarrow 0$ for $\sigma/kT=4$ as a function of $\delta \beta$ at different values of $\beta$. Symbols: Monte Carlo simulation data obtained for random cites at $\beta^{-1} = 0.1$ (black),  $\beta^{-1} = 0.18$ (blue),  $\beta^{-1} = 0.22$ (green), and $\beta^{-1} = 0.3$ (red). The orange dashed line is the result of EGDM, i.e., the data for a cubic lattice at $\beta^{-1} = 0.1$.    }
	\label{fig:mu_vs_F_RS_abd_EGDM}
\end{figure}

In Fig.~\ref{fig:mu_vs_F_RS_abd_EGDM}, we depict by symbols the results of Monte Carlo simulations~\cite{Nenashev2017short,Nenashev2017RANDOMLATTICE} for the system of random sites at $0.1 \leq \beta^{-1} \leq 0.3$ along with the result of EGDM (dashed line), which fits Monte Carlo simulations~\cite{Pasveer2005} on a cubic lattice at $\beta^{-1} = 0.1$. The case $\sigma/kT = 4$ is considered as typical for ODSs at room temperature, since $\sigma$ is usually estimated as $\sigma \simeq 0.1$ eV \cite{Pasveer2005, Lee2021}.

Apparently, $\mu(F)$ on the lattice unequals $\mu(F)$ in a system of random sites. Equation (\ref{eq:ElectrFieldPasveer}) fits the simulation data on a cubic lattice at $\beta^{-1} = 0.1$, though it does not fit the simulation data in a spatially disordered system at $\beta^{-1} = 0.1$ (black circles). The physical mechanism responsible for the drastic difference in $\mu(F)$ between lattices and random sites has been discussed in detail elsewhere~\cite{Nenashev2017RANDOMLATTICE}.

From Fig.~\ref{fig:mu_vs_F_RS_abd_EGDM} we, however, learn that the EGDM, though desired to fit $\mu(F)$ on a cubic lattice at $\beta^{-1} = 0.1$, occasionally fits well $\mu(F)$ on random sites at $\beta^{-1} = 0.18$.
While the value $\beta = N^{-1/3}/a =10$ was standardised in the EGDM~\cite{Coehoorn_Bobbert_Feature2012}, the previously used value $\beta = N^{-1/3}/a = 5$  looks more relevant to ODSs~\cite{Bassler1993}. This conclusion is supported by the estimates for $a$~\cite{Gill1972,Rubel2004,Pasveer2005,Lee2021}, $0.1$ nm $\leq a \leq 0.75$ nm and estimates for $N$ between $N \simeq 1.7 \times 10^{20}$ cm$^{-3}$ and $N \simeq 4.6 \times 10^{21}$ cm$^{-3}$ depending on the material \cite{Bassler1993,Pasveer2005,Lee2021}. The data in Fig.~\ref{fig:mu_vs_F_RS_abd_EGDM} suggest, therefore, that EGDM could be applied to ODSs with a realistic value $\beta^{-1} = a/ N^{-1/3} \simeq 0.18$.
For materials with $\beta^{-1} > 0.18$, one should, instead, use the APAE for $\mu(F)$, as illustrated in Fig.~\ref{fig:mu_vs_F_RS_abd_EGDM}.

Furthermore, EGDM occasionally appears capable to account, in some cases, for the dependece $\mu(T)$ on a system with spatial disorder. In fact, the localization length $a$ affects not only the field dependence $\mu(F)$, but also the temperature dependence $\mu(T)$ \cite{Baranovskii2000,Baranovskii2002a,Lee2021,Sun2021}. This effect is neglected in the EGDM. The effect of $a$ on the dependence $\mu(T)$ can be taken into account, for instance,\cite{Baranovskii2000,Baranovskii2002a,Lee2021,Sun2021} by replacing the constant $\mathfrak{c}_2$ in Eq.~(\ref{eq:Pasveer_T_n_part_b}) by the appropriate functions of the parameter $\beta = N^{-1/3}/a$. The illustration is provided by Fig.~\ref{fig:mu_low_n}, where the slopes of the straight lines depend on $\beta = N^{-1/3}/a$.

At low carrier concentrations and low electric fields, EGDM describes the dependence $\mu(T)$ by Eq.~(\ref{eq:mu_T_Bassler}) with $C = 0.42$.  As evident in Fig.~\ref{fig:mu_low_n}, the dependence $\mu(T)$ for carrier mobility on random sites with $\beta \simeq 5$, typical for ODSs~\cite{Bassler1993}, is described by Eq.~(\ref{eq:mu_T_Bassler}) with $C \simeq 0.40$, which is very close to $C \simeq 0.42$ given by the EGDM. At $\beta$ values different to $\beta \simeq 5$, the APAE should be preferred.

\section{Discussion}
\label{sec:discussion}

\subsection{Comparison with recent results in the literature}
\label{sec:comparison_upretti_et_al}

Several recent studies~\cite{Kemerink2019,Lee2021,Sun2021} were dedicated to the appropriate theoretical description of the dependences $\mu(T,n,F)$ suitable for incorporation into device simulation algorithms.

Upreti \textit{et al}.~\cite{Kemerink2019} considered an analytical theory for $\mu(T,n,F)$ similar to that described in Sec.~\ref{sec:analytical_description} and carried out Monte Carlo simulations used to calibrate parameters in the analytical theory. Their theory was further implemented into the drift-diffusion solver. Along with the theoretical development, Upreti \textit{et al}.~\cite{Kemerink2019} fabricated hole-only and electron-only organic devices and performed measurements of temperature-dependent space-charge-limited currents. It was found that the suggested theory can adequately describe electron
and hole transport in a wide variety of organic semiconductor blends, which are used as the active layer in typical bulk heterojunction organic solar cells~\cite{Deibel_2010}. It was also recognised that EGDM fails to produce an acceptable fit for experimental data obtained on the electron-only devices~\cite{Kemerink2019}. Upreti \textit{et al}. supposed that the reason for the failure of the EGDM for electron-only devices is a small value of parameter $\beta = N^{-1/3}/a =2$, which favors the variable-range hopping (VRH), while EGDM with the large fixed parameter $\beta = 10$ presumably favors the nearest-neighbor hopping mechanism. However, Monte Carlo simulations have proven~\cite{Nenashev2017RANDOMLATTICE} that EGDM is, in fact, based on the VRH transport regime, though on a cubic lattice without spatial disorder. Therefore, more study would be needed to clarify why the EGDM fails to account for experimental data in electron-only devices, though being capable to describe the data in hole-only devices reported by Upreti \textit{et al}.~\cite{Kemerink2019}.

Lee \textit{et al}.~\cite{Lee2021} also addressed an analytical theory for $\mu(T,n,F)$ described in Sec.~\ref{sec:analytical_description}. They implemented
the results in a technology computer-aided design (TCAD) simulation tool, ATLAS, from Silvaco (Silvaco Inc., Atlas Ver. 5.30.0.R (2020)) and deduced parameters of the GDM by comparison with experimental data for ODSs with high carrier mobilities. Only the case of low electric fields was considered by Lee \textit{et al}.~\cite{Lee2021}, who highlighted that the GDM with spatial and energy disorder is superior to the EGDM that lacks spatial disorder and uses only one fixed value $\beta = N^{-1/3}/a =10$ on a simple cubic lattice.

Neither Upreti \textit{et al}.~\cite{Kemerink2019}, nor Lee \textit{et al}.~\cite{Lee2021} attempted to replace the set of analytical equations by a single closed-form expression for $\mu(T,n,F)$ as, for example, APAE derived in Sec.~\ref{sec:APAE}.

Sun \textit{et al}.~\cite{Sun2021} improved the EGDM, though not going beyond the model based on a cubic lattice. One improvement is the replacement of constants $\mathfrak{c}_1$ and $\mathfrak{c}_2$ in Eq.~(\ref{eq:Pasveer_T_n_part_b}) by functions of the parameter $\beta = N^{-1/3}/a$. The other improvement is the replacement of the lattice constant $b$ in Eq.~(\ref{eq:ElectrFieldPasveer}) by $b=10a$ for using the localization length $a$ as adjustable parameter~\cite{Sun2021}. Sun \textit{et al}.~\cite{Sun2021}  solved analytically the degenerate
drift-diffusion equation and extracted model parameters for several organic diodes by comparison with experimental data.

\begin{figure}
	\includegraphics[width=\linewidth]{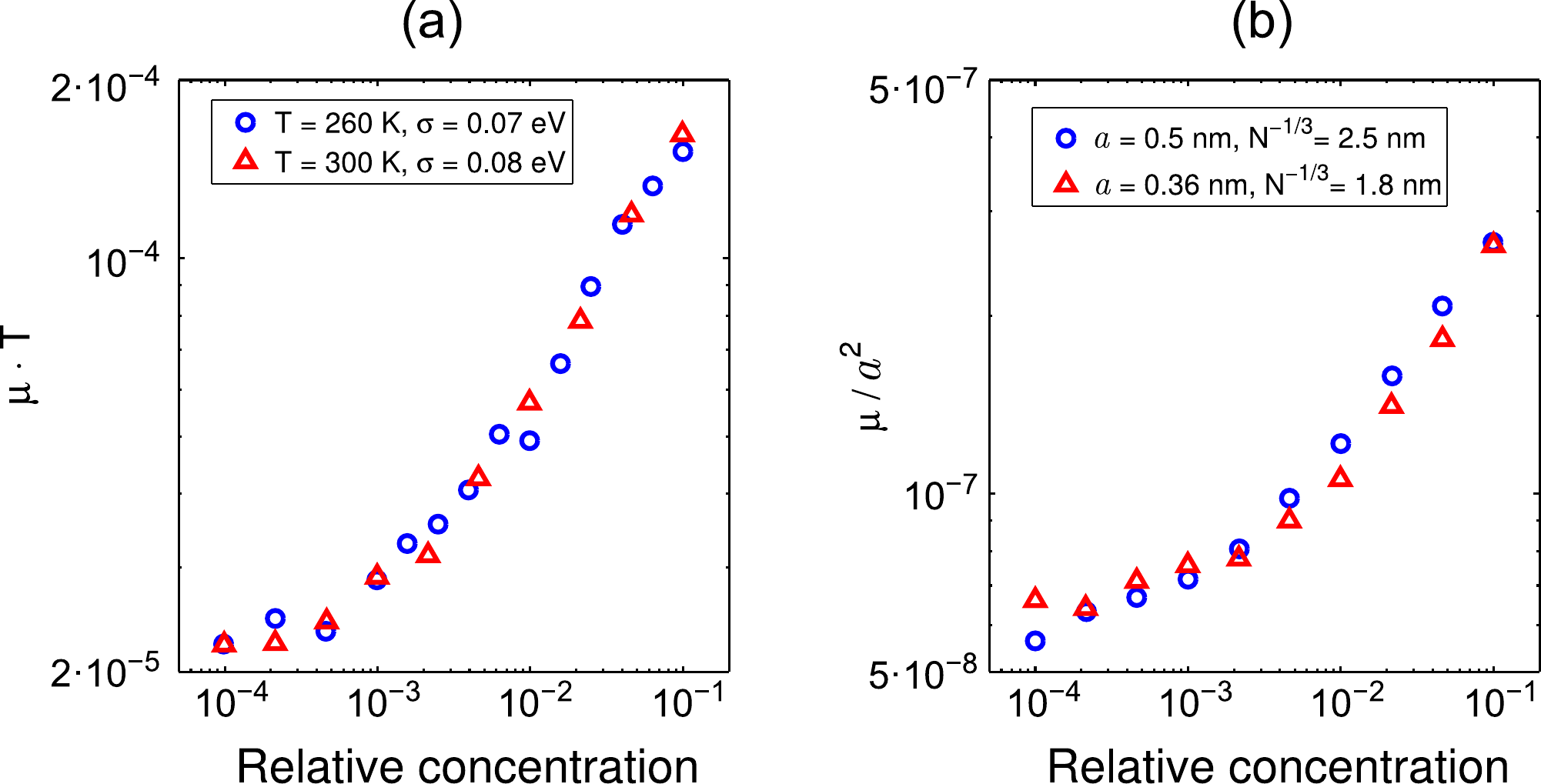}
	\caption{Proof that $\alpha = kT/\sigma$ and $\beta = N^{-1/3}/a$ are the cricial parameters of the GDM and not separately $kT$, $\sigma$, $N^{-1/3}$, $a$. (a) parameter $\alpha = kT/\sigma$ has the same value $\alpha \simeq 0.32$ for the both sets of data; (b) parameter $\beta = N^{-1/3}/a$ has the same value $\beta =5$ for the both sets of data.}
	\label{fig:Upreti_combined}
\end{figure}

These studies have not revealed the combinations of parameters given in Eq.~(\ref{eq:alpha_betta_gamma_delta}) as parametrization of the theoretical model. Instead, parameters $N^{-1/3}$ and $a$ and parameters $\sigma$ and $kT$ were treated separately from each other. In order to emphasize once again the validity of the parametrization given in Sec.~\ref{sec:parametrizzation}, we use in Fig.~\ref{fig:Upreti_combined} the data of Upreti \textit{et al}.~\cite{Kemerink2019} obtained by Monte Carlo simulations for the dependence $\mu(n)$ at different values of parameters $N^{-1/3}$, $a$, $\sigma$, and $kT$. While Upreti \textit{et al}.~\cite{Kemerink2019} plotted their numerical data for $\mu(n)$ as functions of $N^{-1/3}$ at fixed $a$ and as functions of $a$ at fixed $N^{-1/3}$, as well as functions of $\sigma$ at fixed $T$ and functions of $T$ at fixed $\sigma$, we have chosen their combinations of $N^{-1/3}$ and $a$, as well as the combinations of $T$ and $\sigma$, which correspond to the same values of the ratios $\alpha = kT/\sigma$ and $\beta = N^{-1/3}/a$. The plots illustrate that, in fact, $\alpha = kT/\sigma$ and $\beta = N^{-1/3}/a$ control $\mu(n)$ and not separately $N^{-1/3}$, $a$, $\sigma$, and $kT$.
This is the reason why the APAE introduced in Sec.~\ref{sec:APAE} is formulated in terms of only three parameters $\alpha$, $\beta$, $\gamma$ determined in Eq.~(\ref{eq:alpha_betta_gamma_delta}).

\subsection{Conclusions}
\label{sec:Conclusions}

The APAE formulated for $\mu(T,n,F)$ in Sec.~\ref{sec:APAE} is well-justified by classical theories of hopping transport. It is calibrated by Monte Carlo simulations in the framework of the GDM. It is user-friendly and can be easily implemented into device-simulation software.


EGDM~\cite{Pasveer2005} desired to fit simulation data for $\mu(T,n,F)$ in the framework of the GDM on a cubic lattice at fixed value $\beta = N^{-1/3}/a = 10$ occasionally agrees with simulation data on a system of random sites, though at $\beta \simeq 5$. For other $\beta$ values than $\beta \simeq 5$, APAE should be preferred.


\begin{acknowledgments}
A.N. thanks the Faculty of Physics of the Philipps University Marburg
for the kind hospitality during his research stay. S.D.B. and K.M. acknowledge financial
support by the Deutsche
Forschungsgemeinschaft (Research Training Group ``TIDE'', RTG2591).
\end{acknowledgments}

\bibliography{APAE}

\end{document}